\documentclass[11pt,twoside]{article}
\usepackage{./asp2014}

\aspSuppressVolSlug
\resetcounters

\begin{document}

\title{Radio Emission from Short Gamma-ray Bursts in the Multi-messenger Era}
\author{Nicole Lloyd-Ronning$^1$ 
\affil{$^1$Los Alamos National Lab, Los Alamos, NM, USA 87545; \email{lloyd-ronning@lanl.gov}}
}

\paperauthor{Nicole Lloyd-Ronning}{lloyd-ronning@lanl.gov}{ORCID_Or_Blank}{Los Alamos National Lab}{Computational Physics Division}{Los Alamos}{NM}{87544}{USA}


\section{Introduction}

 The most robust model for short gamma-ray bursts (sGRBs) is the merger of two compact objects, such as two neutrons stars (NS-NS) or a neutron star and a black hole (NS-BH).  The timescales and energetics involved in the merger have always made these systems  plausible progenitors for sGRBs \citep{Eich89, Nar92}, but other clues including the location of these bursts in their host galaxies, the lack of associated supernovae, and the observed sGRB rates have provided convincing evidence that these bursts are associated with the older stellar populations expected of compact objects \citep{RRR03, Fox05, Sod06a, LRR07, Berg09, Koc10, LB10, Fong10, Berg10, Fong13, Fong14}. The recent detection of gravitational waves from a neutron star merger \citep{Ab17} and the associated short gamma-ray burst \citep{Ab17b} have provided the smoking-gun evidence that at least some sGRB progenitors result from binary neutron star mergers.

   There has been a concerted effort to follow up short GRBs with the goal of detecting the afterglow and potentially learning more about this class of gamma-ray bursts (for a review, see \citet{Berg14}).  To date, about $93\%, 84\%,$, and $58\%$ of sGRBs have been followed up in the X-ray, optical, and radio respectively \citep{Fong2015}. Of these follow-up efforts, $74\%$ have an X-ray afterglow, $34 \%$ have been seen in the optical, and only $7\%$ in the radio. 
   
  Radio emission in particular has been suggested as a useful tool to help elucidate sGRBs \citep{NP11, MB12, Berg14, Hoto16, Res17}.  In the classic fireball model \citep{MR97,SPN98}, the afterglow emission comes from synchrotron radiation from the forward (and sometimes reverse) shock of the blast wave; the radio component tends to peak at later times, farther from the central engine, more clearly probing the circumstellar environment of the progenitor (compared to the optical and X-ray occurring at earlier times, closer to the cataclysmic event).  In addition, the radio afterglow can in principle be detected when the jet has decelerated (so that the radiation is not strongly beamed), and be detected even when the observer is not aligned with the axis of the GRB jet (as a so-called orphan afterglow; \citet{Lev02,TP02}), or potentially (if there is enough baryon contamination) from a hot cocoon surrounding the merger site \citep{RR02, MB14, Laz16}. Finally, models of compact object mergers predict a significant amount of mass tidally ejected during the merger process \citep{Ruf97, Ros05, OJM07, korobkin12} which can radiate in several ways, including via shocks with the external medium \citep{NP11}.  These latter two emission scenarios are particularly exciting as they are potential electromagnetic counterparts (and hence carry additional important information) to a gravitational wave signal from a double neutron star (DNS) merger event.

\section{Radio Emission Components from Compact Object Mergers}
\label{sec:radio}

 In the standard picture of a relativistic external blast wave, the onset of the afterglow occurs around the deceleration time - i.e. when the blast wave has swept up enough external material to begin to decelerate $t_{dec} \propto (E/n)^{1/3}\Gamma^{-8/3}$ \citep{BM76}.  Assuming the emission mechanism is primarily synchrotron radiation one can calculate the characteristic synchrotron break frequencies at this time, depending on the global and microphysical parameters of the burst. These characteristic frequencies are $\nu_{a}$, the synchrotron self-absorption frequency, $\nu_{m}$, the frequency corresponding to the ``minimum'' (characteristic) electron energy, and $\nu_{c}$, the frequency corresponding to the energy at which an electron loses most of its energy to radiation 
 These expressions are given in Table 2 of \cite{GS02} for both a constant density and wind medium.  
  
In general, $\nu_{a} < \nu_{m} < \nu_{c}$ for the forward shock component. We can - in the context of this model - calculate how the frequencies evolve with time and when they enter the radio band. For optically thin emission, $\nu_{m}$ is usually the most relevant - i.e. the flux is brightest at this frequency and it should dominate the spectrum in the radio band (here, we focus on optically thin emission keeping in mind this provides upper limits to what we should detect in the radio band, although see the brief discussion of the effects of self-absorption below). The cooling frequency $\nu_{c}$ generally stays well above the radio band for an extended time (this may in any case be an oversimplification of the acceleration and cooling processes - see e.g. \citet{LP00} for a discussion on how realistic continual acceleration of particles eliminates this characteristic frequency). For the reverse shock, the minimum electron frequency is roughly $\nu_{m,RS} \approx \nu_{m}/\Gamma^{2}$ (assuming the fraction of energy in the magnetic field is roughly the same for the forward and reverse shock, as explained below), and thus peaks in the radio at earlier times than the forward shock.  We discuss the radio emission from the different components of a compact object merger below.

   
\subsection{Jet Forward Shock}
\label{sec:fs} 
  For a constant density medium, we consider the peak flux at $\nu_{m}$ when it enters the radio band.  This frequency is given by:
  \begin{equation}
  \begin{aligned}
  \nu_{m} = & \ 10^{15} Hz \cdot 
  \left\lbrace 3.73 (p-0.67)(1+z)^{1/2}\epsilon_{e}^{2}\epsilon_{B}^{1/2}E_{52}^{1/2}t_{days}^{-3/2} \right\rbrace
  \label{eq:num}
  \end{aligned}
  \end{equation}
where $p$ is the electron energy power-law index, $z$ is the redshift, $\epsilon_{e}$ is the fraction of energy in the electrons, $\epsilon_{B}$ is the fraction of energy in the magnetic field,  $E_{52}$ is the isotropic equivalent energy normalized to $10^{52}$ ergs and $t_{days}$ is the time normalized to a day. Notice this expression is particularly sensitive to the fraction of energy in the electrons. 
  
  The peak flux at this frequency is given by:
\begin{equation}
\begin{aligned}
       f_{p}(\nu_{m}) = & \ 9930\mu Jy \cdot 
 \left\lbrace (p+0.14)(1+z)\epsilon_{B}^{1/2}n^{1/2}E_{52}d_{L28}^{-2} \right\rbrace
	\label{eq:fpfs}
    \end{aligned}
\end{equation}

\subsection{Jet Reverse Shock}
\label{sec:rs}

  There have been many studies of the reverse shock from a relativistic blast wave (e.g., \citet{MR97, SP99, Kob00, ZKM03, KZ03, ZWD05} and references therein), and the early-time radio flare observation of GRB 990123 has been attributed to the reverse shock \citep{KFS99, NP05}. In addition, \citet{SRR03} examined the expected strength of the reverse shock in six long GRBs, and were able to constrain the hydrodynamic evolution and bulk Lorentz factors of these bursts from this component. 
  
  As pointed out by these references and others, the evolution of the flux and break frequencies in the reverse shock depends on whether the blast wave is Newtonian or relativistic (among other factors), which in turn is related to the shell thickness $\Delta$ estimated from the observed prompt gamma-ray burst duration $T$ by $\Delta \sim cT/(1+z)$. For a thick shell, $\Delta > l/2\Gamma^{8/3}$, where $l$ is the Sedov length in an ISM medium $ \equiv (3E/4\pi n m_{p}c^{2})^{1/3}$, the reverse shock has time to become relativistic and the standard Blandford-McKee solution applies.  For a thin shell, the reverse shock remains Newtonian and the Lorentz factor of this shock evolves as $\Gamma_{RS} \sim r^{-g}$, with $g \sim 2$ \citep{Kob00}.  Short bursts with $T< 1s$ are likely in the thin shell - and therefore Newtonian - regime.  However, we note that for a range of $g$ values, the time evolution of the flux and characteristic frequencies are fairly similar between the relativistic and Newtonian regimes.  
  
   
 Generally speaking, because of the higher mass density in the shell, the peak flux in the reverse shock $f_{p, RS}$ will be higher by a factor of $\Gamma$ relative to the forward shock, 
  \begin{equation}
  f_{p, RS} \approx \Gamma f_{p, FS}
  \end{equation}
  but the minimum electron frequency in the reverse shock $\nu_{m, RS}$ will be lower by a factor of $\Gamma^{2}$,
  \begin{equation} 
  \nu_{m, RS} \approx \nu_{m, FS}/\Gamma^{2}
  \end{equation}
  assuming the forward and reverse shock have the same fraction of energy in the magnetic field (not necessarily a well-justified assumption). 
  
 \subsubsection{Self-absorbed Reverse Shock}
 
  At radio wavelengths, synchrotron self-absorption should be considered - under certain conditions lower energy photons are self-absorbed, and the flux is suppressed.  Self-absorption may be particularly relevant in the region of reverse shock, where the density is higher relative to the forward shock region. \citet{RZ16} calculated the relevance of the self-absorption frequency and flux in the reverse shock, before and after shock crossing. For later time radio emission, we consider the frequencies and fluxes after the shock crosses the thin shell (but see their Appendix A.1 for expressions in all ranges of parameter space).  

\noindent Roughly, at the high radio frequencies we are considering here, the flux at the time of the peak can be obtained from equation 30 of \citet{RZ16}:
\begin{equation}
f_{p,RS} = f_{p, RS, \nu_{m}}(\nu_{a, RS}/\nu_{m, RS})^{-\beta}
\end{equation}
\noindent where $\beta=(p-1)/2$.\\
The reverse shock flux is suppressed at a minimum by factors ranging from about 0.3 to 0.01. We emphasize again, therefore, that our estimates are upper limits to the emission from the reverse shock.

\subsection{Jet Off-axis Emission}
\label{sec:offax}
   Because GRB outflows are relativistic, off-axis emission is highly suppressed due to the relativistic beaming of the radiation.  However, once the Lorentz factor decreases to a value on the order of the inverse of the viewing angle of the observer, $\Gamma$ $\sim 1/\theta_{v}$, the flux is similar to an on-axis observer (\citet{Rhds97, Gran02};  See also Figures 7 and 8 of \citet{GvdH14}).


   As a simple estimate of the peak of the off-axis emission, we calculate the flux at a time when the blast wave is no longer relativistic, so that the radiation is no longer relativistically beamed and is detectable to an observer well off-axis. This time - which occurs after the jet break time - is also an upper limit to the time of detection of the unbeamed flux.   The blast wave becomes non-relativistic around $\sim 1 yr (E_{52}/n)^{1/3}$ \citep{Pir04}; at this time, $\nu_{m}$ is usually well below the typical observed radio band.  Hence, to estimate the radio emission we need the flux above $\nu_{m}$ at a time when the blast wave has decelerated enough to become non-relativistic.  Using \cite{FWK00}, the flux at 8.46GHz at the non-relativistic transition is:
  \begin{equation}
  \label{offax}   
  \begin{aligned}
  F_{\nu>\nu_{m}} = & \ 1635\mu Jy \lbrace (1+z)\epsilon_{e}\epsilon_{B}^{3/4}n^{3/4}E_{52} (t/t_{NR})^{3(1-p)/2 + 3/5} \\ & d_{L28}^{-2} (\nu/8.46GHz)^{(1-p)/2}(\theta_{j}/0.1)^{2}\rbrace 
  \end{aligned}  
  \end{equation}

\noindent where $\theta_{j}$ is the physical opening angle of the jet.  Note there are many ways to model off-axis emission (see, e.g. \citet{NPG02}, \citet{Sod06}, \citet{Wax04}, \citet{Oren04}), depending on the underlying assumptions of the behavior of the jet. This emission - the magnitude and time of the of the peak flux in particular - depends of course on the observer viewing angle (for a recent estimation of radio flux dependence on viewing angle, see \cite{CC18}).  We emphasize again that we have taken the very conservative estimate of the flux at a time when the blast wave is non-relativistic, and the emission is isotropic and observable to viewers at all angles.

  \subsection{Tidal or Dynamical Ejecta}
\label{sec:tidej}

In addition to emission from the jet component, there is also emission from mass either dynamically ejected or from winds during the merger process \citep{Ros99, RJ01, YST08, Rez10}.   This mass (typically in the range from $\sim 0.01 - .1M_{\odot}$) is
ejected with a velocity $\beta_{i} = v/c \sim 0.1$ \citep{NP11}, and can shock with the external medium, emitting in the radio band. We note that some simulations have found higher values for the tidal ejecta velocity, $\sim 0.2c-0.3c$, (e.g. \citet{Rad16}), which will increase the values of the peak flux; however, this value depends strongly on the equation of state of the neutron star, and a definitive value for the ejecta velocity from neutron star mergers is far from settled (for a brief discussion of this issue, see \cite{Met17}). We note importantly that the bulk of the ejecta mass (which likely came from the wind component of the merger) from GW170817 was fit with a velocity of $0.08c$ \citet{Troja17}.

If the observed frequency is above the minimum electron frequency $\nu_{m}$ and the self-absorption frequency $\nu_{a}$ - both expected to be less than around 1GHz for canonical values of energies and densities for mergers - then the flux will peak at a time \citep{NP11, NPR13}:

\begin{equation}
\label{eq: tdectidal}
t_{dec} \approx 300 days E_{51}^{1/3}n^{-1/3}\beta_{i}^{-5/3}
\end{equation}

Where $t_{dec} = R_{dec}/c\beta_{i}$ is the deceleration time (when the ejecta has swept up mass comparable to its own at a radius $R_{dec}$; after $t_{dec}$, the flow decelerates in Sedov-Taylor flow $\beta \approx \beta_{i}(R/R_{dec})^{-3/2}$).

 The peak of the specific flux occurs at $t_{dec}$ and is given by \citep{NP11}:

\begin{equation}
\label{eq:fluxtidal}
\begin{aligned}
F_{\nu > \nu_{m}} \approx 4012\mu & \ Jy E_{52}n^{\frac{p+1}{4}}\epsilon_{B}^{\frac{p+1}{4}} \epsilon_{e}^{p-1}  \beta_{i}^{\frac{5p-7}{2}} d_{28}^{-2} (\frac{\nu_{obs}}{8.46GHz})^{-\frac{p-1}{2}}
\end{aligned}
\end{equation}

Note our change of units compared to \citet{NP11} and \citet{Berg14}, that  $\epsilon_{B}$ and $\epsilon_{e}$ are the absolute, non-normalized values, and $\nu_{obs}$ is the observed frequency normalized to 8.46 GHz.

 \section{Predicted Radio Emission from a Compact Merger}
 \label{sec:rademit}
\begin{figure*}
\includegraphics[width=5.5in]{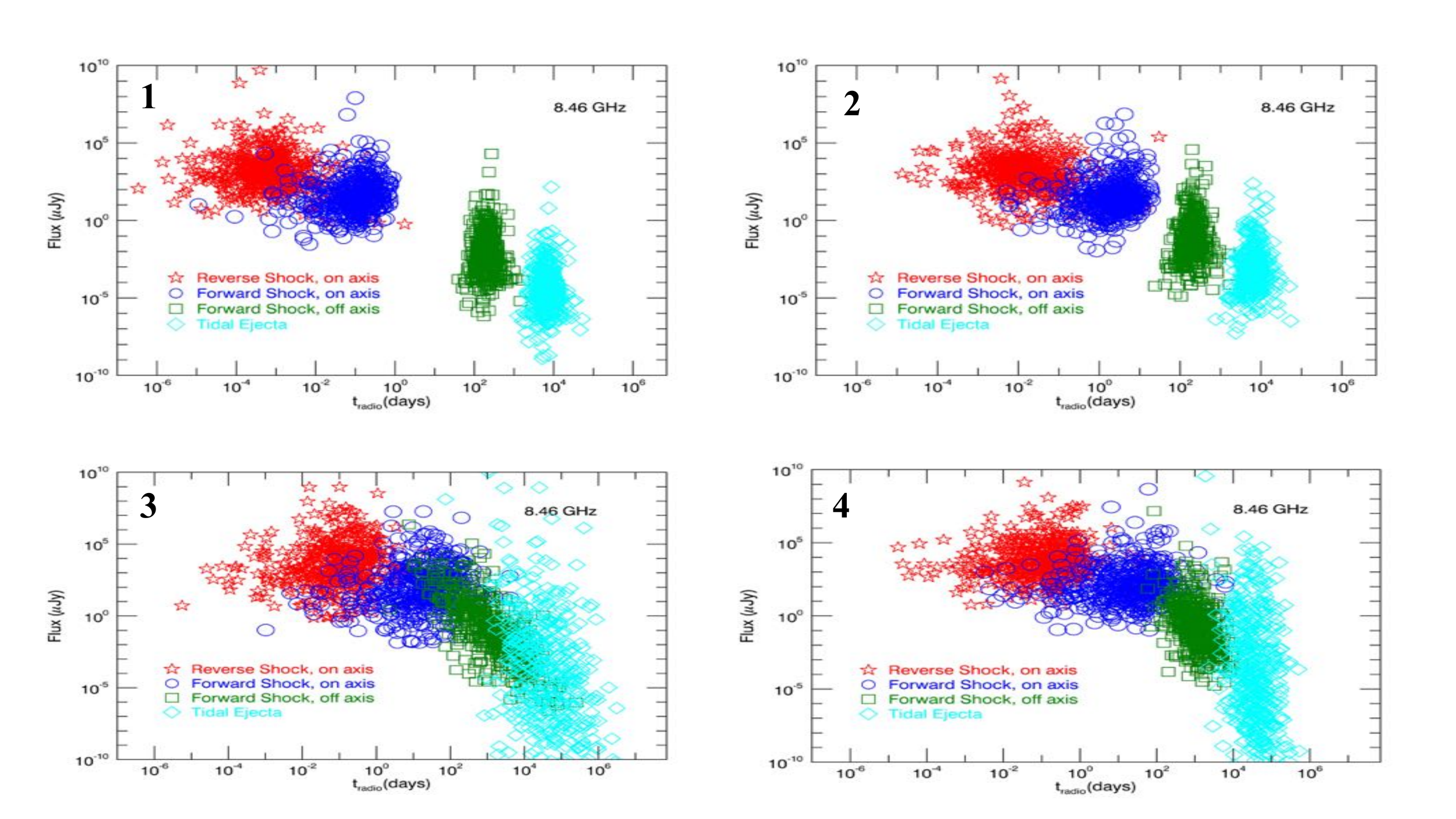}
\caption{{\bf Top Left, Model 1:} Flux as a function of time using densities from a distribution centered around 0.1 cm$^{-3}$, a mean energy of $E=10^{50}erg$, a mean Lorentz factor of $50$, a mean redshift of $0.3$, a fixed value of $p=2.3$, and  $\epsilon_{e}$ and $\epsilon_{B}$ drawn from a Gaussian with a mean of 0.01. For the tidal ejecta component, we assumed an ejecta velocity $\beta_{i} = 0.1$. {\bf Top Right, Model 2:} Same as Model 1 in left panel, but with $\epsilon_{e}$ and $\epsilon_{B}$ drawn from a Gaussian with a mean of 0.1 and 0.01 respectively. {\bf Bottom Left, Model 3:} Flux as a function of time using distributions of parameters mimicking those from the multi-wavelength fits of \citet{Fong2015}.  Densities are from a distribution centered around 0.005 cm$^{-3}$ with a large spread ($\sim 2.5$ orders of magnitude), a mean energy of $E = 10^{51}erg$, a mean redshift of $0.5$, a mean value of $p=2.43$. The parameters  $\epsilon_{e}$ and $\epsilon_{B}$ are both drawn from a Gaussian with a mean of 0.1.  The Lorentz factor employed is $\Gamma = 50$. {\bf Bottom Right, Model 4:} Same as Model 3, but with less spread in the density distribution (a standard deviation of 1.25 orders of magnitude as opposed to 2.5 orders of magnitude).}
\label{fig:Radfluxall}
\end{figure*}

  Figure~\ref{fig:Radfluxall} shows the predicted radio flux at 8.46 GHz from the forward shock (blue circles), reverse shock (red stars), off-axis emission (green squares), and tidal ejecta (light blue diamonds) for different models, described in the figure caption.   For the top panels of Figure~\ref{fig:Radfluxall}, we drew each of the physical GRB parameters from a Gaussian distribution with a normalized mean of 1, except for the electron energy power-law index which was fixed at $p=2.3$, and - in the case of the tidal ejecta component - an ejecta velocity of $\beta_{i}=0.1$. We used a mean density of  0.1 cm$^{-3}$, a mean energy of $E=10^{50}erg$, a mean Lorentz factor of $50$, and a mean redshift of $0.3$.  The averages of the parameters $\epsilon_{e}$ and $\epsilon_{B}$ varied from 0.1 to 0.01.  These plots show that the variation in GRB parameters can give a wide range of flux values.

  For the bottom panels in Figure ~\ref{fig:Radfluxall}, we plot the predicted radio fluxes, where spectral parameters were taken from distributions that mimic the observed or fitted distributions from \citet{Fong2015}. The difference between these two bottom panels is the choice of standard deviation of the density distribution in log space ($2.5$ and $1.5$ for the left and right panels respectively).  Again, there is a large distribution in flux from the various emission components for our choice of spectral parameters. In principle, we may be able to distinguish these components from the time at which they peak in the radio.  This is relatively well delineated, with the reverse shock peaking earliest, then the forward shock, off-axis emission and finally, the dynamical ejecta component.   As is always the case with gamma-ray bursts, the more multi-band temporal and spectral information we can obtain for each burst, the better chance we have of putting constraints on their emission and therefore underlying spectral parameters. In particular, multi-wavelength observations will help us constrain the characteristic frequencies (and their corresponding fluxes) of the spectrum, helping to break some of the degeneracy between various physical GRB parameters.  
 In addition to this, however, there is a real need from a theoretical standpoint to better understand the microphysics of relativistic shocks (through, for example, long timescale particle-in-cell simulations) and place tighter constraints on the electron energy index $p$, as well as the parameters $\epsilon_{e}$ and $\epsilon_{B}$, both of which can in principle take on a wide range of values (spanning orders of magnitude) and contribute to large uncertainty in the expected flux.   
  

\subsection{Prospects for ''Orphan" Components, and EM Detections Coincident with aLIGO}
  Because the off-axis and tidal ejecta components are quasi-isotropic, we can ask what fraction of these components we might detect in the radio, if the GRB jet is not directed toward us (such that the prompt gamma-ray signal is not detected).  With the successful detection of gravitational waves from astrophysical events \citep{Ab16}, and in particular the detection of gravitational waves and electromagnetic emission from GW170817 \citet{Ab17}, we have definitively entered a new era of multi-messenger astronomy. 
  
\begin{figure*}[t]
\centering
\includegraphics[width=5.in]{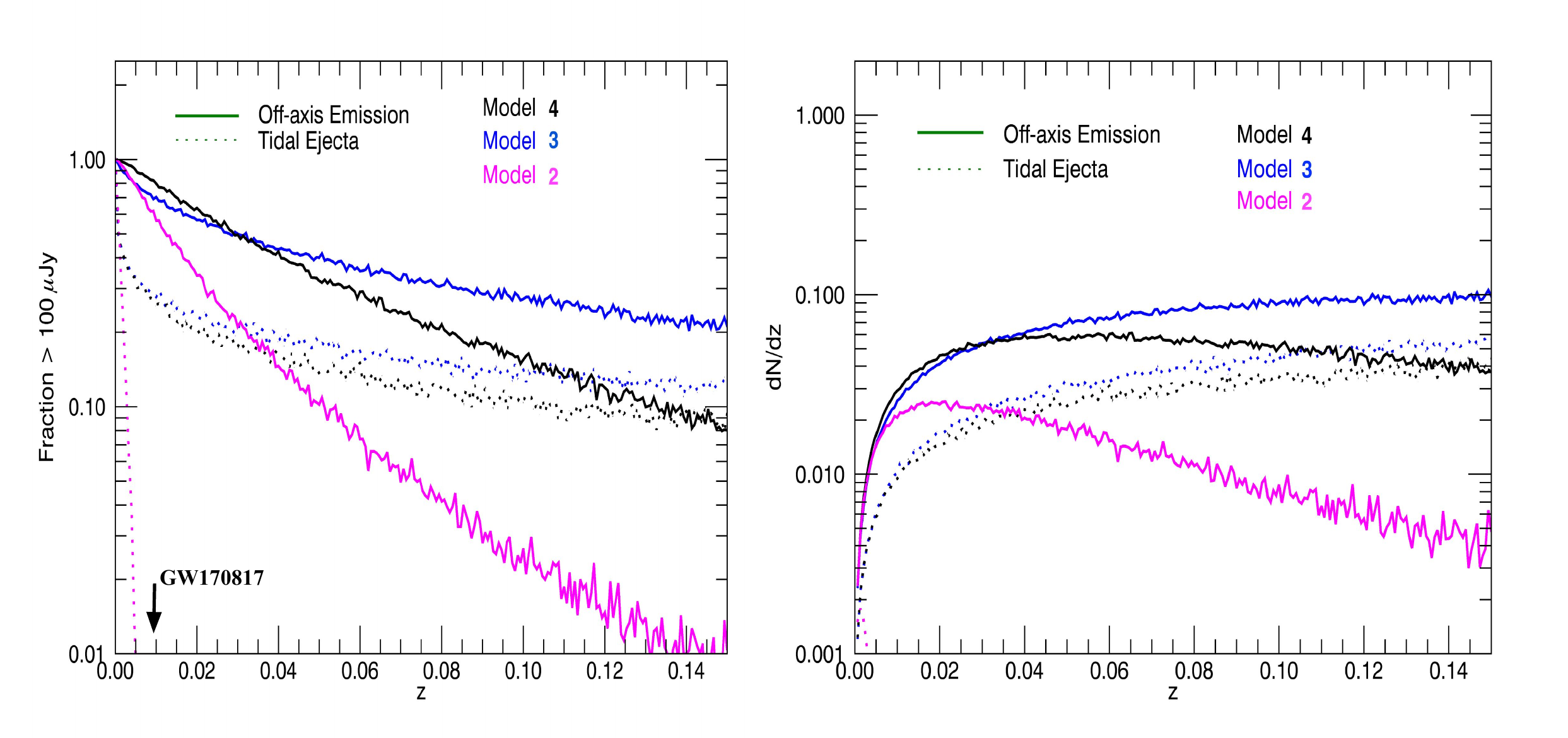}
\caption{{\bf Left panel:} Fraction of GRBs with off-axis emission (solid lines) and tidal ejecta emission (dotted lines) above $100 \mu Jy$ at low redshifts (relevant to the sensitivity of aLIGO in detecting gravitational waves from neutron star mergers), for models 2 - 4 described in Figure~\ref{fig:Radfluxall}).  {\bf Right panel:} The fraction from the left panel multiplied by the sGRB differential distribution as a function of z.  Details on the distribution used are described in the text.  Note we have marked the distance of the neutron star merger/gamma-ray burst GW170817 for reference.}
\label{fig:fracrateLIGO}
\end{figure*}

\begin{figure*}
\centering
\includegraphics[width=5.0in]{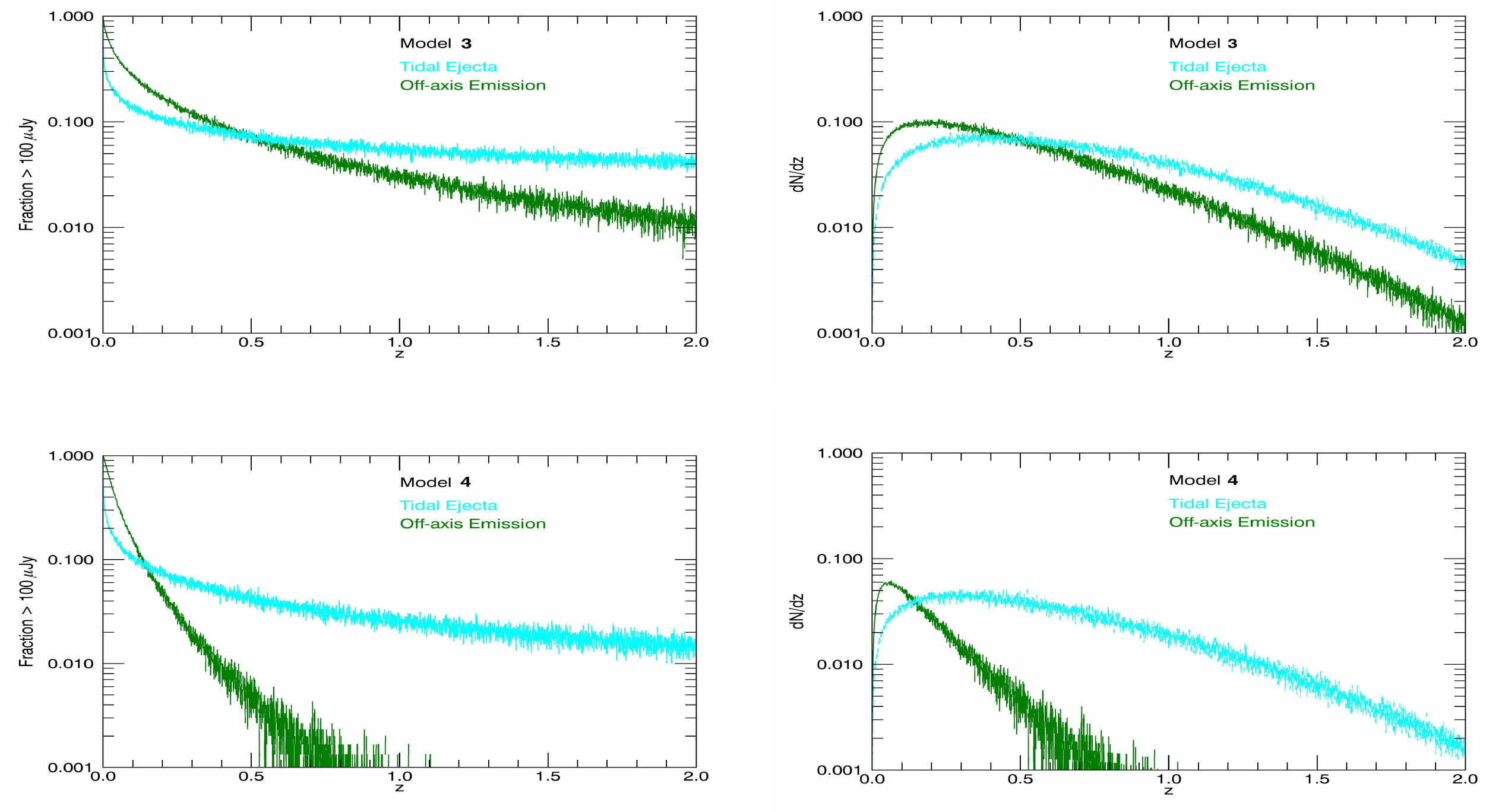}
\caption{{\bf Left panels:} Fraction of GRBs with off-axis emission (green) and tidal ejecta emission (cyan) above $100 \mu Jy$ as a function of redshift, for the Models 3 (top) and 4 (bottom) described in Figure ~\ref{fig:Radfluxall}.  {\bf Right panels:} The fraction from the left panels multiplied by the sGRB differential distribution as a function of z.  Details on the distribution used are described in the text.}
\label{fig:fracrate}
\end{figure*}

  The left panel of Figure~\ref{fig:fracrateLIGO} shows the fraction of GRBs with tidal ejecta (dotted lines) and off-axis components (solid lines) that fall above a conservative limit of $100 \mu Jy$ out to a redshift $z \sim .15$ (relevant to aLIGO's design sensitivity in detecting a GW signal from a DNS). We show curves for models 2, 3 and 4 in Figure~\ref{fig:Radfluxall}, under the assumption that we have full time coverage (i.e. that we are pointing our telescopes at the GRB at the time of the peak).  We find that for the models that mimic the observed/fitted sGRB parameters from \citet{Fong2015}, we expect to detect the quasi-isotropic radio emission component in $\sim 10 - 20 \%$ of events that occur at distances to which aLIGO will be (at design specification) sensitive to NS mergers ($\sim 100 Mpc$). Note that at the distance of GW170817 (40 Mpc or z=0.009783), we expect to detect the off-axis radio component from the sGRB jet and it appears that this component was indeed detected \citep{Alex17b}.
  The right panel of Figure~\ref{fig:fracrateLIGO} shows this fraction multiplied by a fiducial sGRB differential redshift distribution, modeled after the solid line in Figure 4 of \citet{GP06} - a distribution of redshift that peaks at about $z=0.5$ and falls off gradually to zero by $z \sim 2$. The left and right panels of Figure~\ref{fig:fracrate} are similar to Figure~\ref{fig:fracrateLIGO}, but extended to a redshift of 2.  
    


\section{The Future of sGRBs with ngVLA}
\label{sec:conc}

The ngVLA - with its broad spectral range and sensitivity - will allow us for the first time to better discriminate between different components of sGRB emission and elucidate the physics behind these fascinating events.

 We can get additional important information on short gamma-ray bursts if there is rapid follow-up ($< 1$ day) in the radio - this will give the best chance of detecting the reverse shock emission component, and the ngVLA may afford us this opportunity. 
 The circumburst density must also be low enough to allow for a slow-cooling reverse shock (as mentioned in \citet{Las13, Las16}), but such densities are expected for compact object binary progenitors of sGRBs.

 The detectable radio components to DNS or NS-BH or mergers are not necessarily easy to distinguish given the observed range of sGRB physical parameters (in other words, the distributions of sGRB parameters based on afterglow fits give a wide spread in their radio flux values).  However, detecting the time of the peak of each emission component is perhaps the biggest distinguishing factor (see Figure~\ref{fig:Radfluxall}), and spectral information constraining the peak flux of each component is crucial. The ngVLA's wider spectral coverage will be crucial in separating these components.

  The fraction of sGRBs with an off-axis ``orphan'' component detectable in the radio (at late times when the blast wave is non-relativistic and the emission is no longer beamed) and potentially coincident with a gravitational wave signal from a DNS merger is $\sim 0.1$ at distances to which aLIGO would be sensitive to such a signal. This statement is of course model dependent. However, for a reasonable set of models based on the \citet{Fong2015} fits to sGRB afterglow data, {\em we expect to detect the quasi-isotropic radio emission for about $10 \%$ (for the tidal ejecta emission) and up to $20 \%$ (for the off-axis emission) of sGRBs at a redshift of $z \sim 0.15$}. For shorter distances (e.g. the distance of GW170817 of 40 Mpc or z=0.009783), there is a very high probability to detect the radio emission from the off-axis jet component for most fiducial models. At larger distances, the detectable fraction of these quasi-isotropic components falls off significantly.


 The radio emission is a key piece of the puzzle in understanding GRB emission and combining ngVLA observations with additional multi-wavelength follow-up will allow us to constrain the underlying physics of the outflow producing short gamma-ray bursts.  Efforts in this vein are particularly timely in light potential additional detections by aLIGO of gravitational wave emission from a double neutron star merger.  A better understanding of the various components of electromagnetic emission from these objects will provide a more complete picture of these systems and ultimately help us understand their role in the context of stellar evolution in the universe. 

\acknowledgements  We thank Dieter Hartmann, Chris Fryer and Brandon Wiggins for discussions related to this work. Work at LANL was done under the auspices of the National Nuclear Security Administration of the US Department of Energy at Los Alamos National Laboratory LA-UR-17-24900 




\end{document}